\documentclass[aps,twocolumn,showpacs,amsmath,amssymb]{revtex4}
\usepackage{graphicx}% Include figure files
\usepackage{dcolumn}% Align table columns on decimal point
\usepackage{bm}% bold math
\begin{document}
\baselineskip 16pt

\title{The
Einstein-Hamilton-Jacobi equation: Searching the classical
solution for barotropic  FRW
 \vspace{-6pt}}

\author{S. R. Berbena, A. V. Berrocal,  and
J. Socorro}
\address{Instituto de F\'{\i}sica de la Universidad de Guanajuato,\\
Apartado Postal E-143, C.P. 37150, Le\'on, Guanajuato,
M\'exico.\\}
\author{ Luis O. Pimentel}
\address{Departamento de F\'{\i}sica, Universidad Aut\'onoma Metropolitana\\
Apartado Postal 55-534, C.P. 09340, M\'exico, D.F.\\}

\begin{abstract}
The dynamical evolution of the scale factor of FRW cosmological
model is presented, when the equation of state of the material
content assume the form $\rm p=\gamma \rho$, $\rm
\gamma=constant$, including the cosmological term. We  use the WKB
approximation and the relation with the Einstein-Hamilton-Jacobi
equation  to obtain the exact solutions.
\end{abstract}

\maketitle

\section{Introduction}
The behaviour of the cosmological scale factor $\rm A(t)$ in
solutions of Einstein's field equations with the
Friedmann-Robertson-Walker line element has been the subject of
numerous studies, where the presentations tend to focus on models
in which $\rm p=0$ and there is no cosmological constant ($\rm
\Lambda=0$). Some treatments include the cosmological constant
\cite{uno,dos,tres,cuatro,cinco} and the pressure p is given in
terms of density $\rho$ by an equation of state $\rm p=p(\rho)$
and $\Lambda\not= 0$, for particular values in the $ \gamma$
parameter \cite{seis,siete,ocho,nueve}.

The standart model of cosmology is based on Einstein's General
Relativity theory, which can be derived from the geometric
Einstein-Hilbert Lagrangian
\begin{equation}
\rm {\cal L}_{geo} = \frac{1}{16 \pi G} \sqrt{-g} \, R,
\end{equation}
where R is the Ricci scalar, G the Newton constant, and $\rm
g=|g_{\mu\nu}|$ the determinant of the metric tensor. By
performing the metric variation of this equation, one obtains the
well known Einstein's field equations
\begin{equation}
\rm R_{\mu\nu} - \frac{1}{2}R \, g_{\mu\nu} = - 8\pi G T_{\mu\nu},
\label{field}
\end{equation}
where $\rm T_{\mu\nu}$ is the energy-momentum stress tensor,
associated with a matter lagrangian, which is the source of
gravitation, assigning the corresponding equation of state, which
varies during different epochs of the history of the universe.

Introducing a symmetry through the metric tensor, in cosmology one
assumes a simple one according to the cosmological principle that
states the universe is both homogeneous and isotropic. This
homogeneous and isotropic space-time symmetry was originally
studied by Friedmann, Robertson, and Walker (FRW). The symmmetry
is encoded in the special form of following line element
\begin{equation}
\rm ds^2= -N^2(t) dt^2 + A^2(t)\left[\frac{dr^2}{1-\kappa r^2}
+r^2 d \Omega^2   \right]
\end{equation}
where $\rm A(t)$ is the scale factor, N(t) the lapse function,
$\kappa$ is the constant curvature, taking the values $0,+1,-1$
(flat, closed and open space, respectively).

The FRW solutions to the Einstein field equation (\ref{field})
represent a cornerstone in the development of modern cosmology,
since with them it is posible to undertand the expansion of
universe.

Recently, Faraoni \cite{faraoni} introduced one procedure based in
the Riccati differential equation, obtained by the combination of
the Einstein field equation, resulting in the same solutions
obtained by the standard procedure
\cite{uno,dos,tres,cuatro,cinco,inin}, without the cosmological
term. This alternative approach is more direct than the standard
one, which  was used in the factorization procedure in the
supersymmetric level \cite{rosu1,rosu2} for to obtain the both
iso-spectral potential and function in particular one dimension
systems.

The set of differential equation for the FRW cosmological model,
including the cosmological term, become
\begin{eqnarray}
\rm \frac{\ddot A}{A} &=& \rm -\frac{4\pi G}{3} (\rho + 3 p)-
\frac{\Lambda}{3}
\label{uno}\\
\rm \left( \frac{\dot A}{A}  \right )^2 &=& \frac{8\pi G}{3} \rho
- \frac{\Lambda}{3} - \frac{\kappa}{A^2}. \label{dos}
\end{eqnarray}
the overdot means $\rm d/dt$.

In the literature one can find the well-known  classical behaviour to
the scale factor for $\kappa=0$ $\rm $\cite{inin}
\begin{equation}
A= \left[ 6\pi G M_\gamma \left(\gamma +1\right)^2
\right]^{\frac{1}{3(\gamma + 1)}} (t-t_0)^{\frac{2}{3(\gamma
+1)}}.
\end{equation}
Taking different values for the constant $\gamma$ we have the
following subcases
\begin{equation}
A= \left\{
\begin{tabular}{lllll}
$\left[ \frac{32}{3}\pi G M_{\frac{1}{3}} \right]^{\frac{1}{4}}
t^{\frac{1}{2}}$&\qquad& for & $\gamma=\frac{1}{3}$ & radiation\\
$\left[ 6\pi G M_0 \right]^{\frac{1}{3}}
t^{\frac{2}{3}}$&\qquad &for & $\gamma=0$ & dust\\
$\left[ 24\pi G M_1 \right]^{\frac{1}{6}} t^{\frac{1}{3}}$&\qquad
&for & $\gamma=1$ & stiff fluid
\end{tabular}
\right.
\end{equation}
However, for the case $\gamma=-1$ the solution becomes to
exponential
\begin{equation}
\rm A= A_0 e^{H t}, \qquad with~H=2\sqrt{\frac{2}{3}\pi G M_{-1}},
\end{equation}
here, we consider the sign (+) in the exponential function,
because we consider the inflationary behaviour. These solutions
will be compared with the solutions found by our method.

The main purpose of this work is the introduction of the WKB-like
procedure for to  calculate the function $\rm A(t)$, including the
function $\Phi(A)$, function that play a important role in the
supersymmetric fashion\cite{socorro1,socorro2}, called the
superpotential function, into the Hamiltonian formalism for to
solve the Einsten-Hamilton-Jacobi equation. Also, we include the
cosmological term in the formalism.

The remainder of the paper is organized as follow. The procedure
that include the Einstein-Hamilton-Jacobi equation and the master
equation is described in Sec. II. In Sec. III we present the
exact solutions for the master equation found for this model,
including the corresponding analysis of them. Finally, the Sec. IV
is devoted to comments.

\section{Einstein-Hamilton-Jacobi equation: The WKB-like method}

We will use the total Lagrangian  for a homogeneous and isotropic
universe (FRW cosmological model), and perfect-fluid like ordinary
matter with pressure $\rm p$ and energy density $\rm \rho$, and
barotropic state equation $\rm p=\gamma \rho$, including the
cosmological term $\Lambda$\cite{Ryan,Pazos}.

\begin{equation}
\rm L= \frac{6 A}{N} \left( \frac{dA}{dt}  \right)^2 - 6 \kappa N
A - 2 N \Lambda A^3 + 16 \pi G N M_\gamma A^{-3\gamma}.
\label{lagrangian}
\end{equation}
We define the canonical momentum conjugate to the generalized
coordinate $\rm A$ (scale factor) as
\begin{equation}
\rm \Pi_A \equiv \frac{\partial L}{ \partial \dot A}= \frac{12
A}{N} \frac{dA}{dt}, \label{momentum}
\end{equation}
 The canonical hamiltonian function has
the following form

\begin{eqnarray}
\rm &&L = \Pi_A \dot A - N {\cal H}= \Pi_A \dot A \nonumber\\
&&- N \left[\frac{\Pi_A^2}{24 A} + 6 \kappa A + 2 \Lambda A^3 - 16
\pi G M_\gamma A^{-3\gamma} \right ], \label{canonical}
\end{eqnarray}
where
\begin{equation}
{\cal H} =\rm \frac{1}{24 A}\left[\Pi_A^2 + 144 \kappa A^2 + 48
\Lambda A^4 - 384 \pi G M_\gamma A^{-3\gamma +1} \right ].
\label{hamiltonian}
\end{equation}
Performing the variation of (\ref{canonical}) with respect to N,
$\rm {\partial L}/{\partial N}= 0$, implies the well-known
result ${\cal H}=0$.

AT this point we can do two things: i) The quantization procedure,
imposing the quantization condition on ${\cal  H} \rightarrow \hat
{\cal H}$, where $\hat {\cal H}$ is an operator, and applying this
hamiltonian operator
 to the wave function $\Psi$, we obtain the Wheeler-DeWitt (WDW) equation
in the minisuperspace
\begin{equation}
\hat {\cal H} \Psi=0,
\end{equation}
and ii) WKB like method, if one perform the transformation
\begin{equation}
\rm \Pi_A = \frac{d\Phi}{dA} \label{ansatz}
\end{equation}
in (\ref{hamiltonian}), becomes the Einstein-Hamilton-Jacobi
equation, when $\Phi$ is the superpotential function that is
related to the physical potential under consideration.

We shall use the part ii) as an alternative method for obtain the
classical solutions to the FRW cosmological model.

Introducing the ansatz (\ref{ansatz}) into the Eq.
(\ref{hamiltonian}) we get

$$%\begin{equation*}
 \rm \left[\left(\frac{d\Phi}{dA} \right)^2 + 144 \kappa A^2
+ 48 \Lambda A^4 - 384 \pi G M_\gamma A^{-3\gamma +1} \right ]=0
$$%\end{equation*}

\begin{equation}
\rm \frac{d\Phi}{dA}=\pm 12 A \sqrt{\frac{8}{3} \pi G M_\gamma
A^{-(3\gamma +1)} -\frac{\Lambda}{3} A^2 - \kappa}  . \label{EHJ}
\end{equation}

Relating the equations (\ref{momentum},\ref{ansatz}) and
(\ref{EHJ}), we
 obtain the classical evolution for the scale factor in term of the
``cosmic time'' $\tau$ defined by $\rm d\tau = N(t) dt$, through
the following master equation

\begin{equation}
\rm d\tau = \frac{dA}{\sqrt{\frac{8}{3} \pi G M_\gamma
A^{-(3\gamma +1)} -\frac{\Lambda}{3} A^2 - \kappa} }.
\label{evolution}
\end{equation}

that correspond to Eq. (\ref{dos}) in the gauge N=1.

This equation is not easy to solve in general way for all values
in the $\gamma$ parameter. However, we can solve this one for
particular values in two sectors in the $\gamma$ parameter:
\begin{enumerate}
\item{} $\gamma < 0$, it it say $(-1/3, -2/3, -1)$ and $\Lambda
\not=0$.
This is the phenomenon commonly known as inflation-like.
\item{} $\gamma =1/3$,  $\Lambda \not=0$, any $\kappa$.
\item{} $\gamma >0$,  $\Lambda =0$.
\end{enumerate}
In the following, we describe its behaviour for the
scale factor.

\section{Solution to the master equation}
Here, we obtain analytic solutions for the scale factor, via the
master equation rewritten in terms of a ``conformal time''
coordinate T. In some cases, will be necessary
to drop the cosmological term, with the end of obtaining the
corresponding exact solution.

\subsection{$\gamma < 0$, \,\, inflation-like phenomenon}
Considering some negative values for the $\gamma$ parameter,
namely,  $\gamma=-1, -1/3, -2/3$, we have
\begin{enumerate}
 \item  $\gamma=-1$, the equation (\ref{evolution}) is  (for simplicity we choose
the changes $\rm A \rightarrow x$, $\rm a_\gamma=\frac{8}{3}\pi G
M_\gamma$, $\rm b=-\Lambda/3$)
\begin{equation}
\rm d\tau = \frac{dx}{\sqrt{\left( a_{-1} +b \right)x^2 - \kappa}}
\label{infla}
\end{equation}
integrating (\ref{infla}) and  inverting, we obtain
\begin{equation}
\rm A(\tau) = \sqrt{ \frac{3\kappa}{\Lambda - 8\pi G M_{-1}}}
 \quad \sinh \left[ \sqrt{\frac{8\pi G M_{-1}}{3}
- \frac{\Lambda}{3} } \tau  \right] .
\end{equation}
The character of  this solution is related to   the cosmological term
$\Lambda$ and the curvature parameter $\rm \kappa$ as follows,
\begin{enumerate}
\item{} For $\rm \kappa=1 $, the behaviour is inflationary.
\item{} For $\rm \kappa =-1$, the behaviour will be inflationary if
$\rm M_{-1} > (\Lambda + 3)/8 \pi G > 0$.
\item{} For $\rm \kappa =0$, we will solve the original equation
(\ref{evolution}), obtaining
\begin{eqnarray}
\rm A(\tau) &=&   \rm m_1 \,\, exp \left[2\sqrt{\frac{2}{3}\pi G
M_{-1} - \frac{\Lambda}{3}}
\,\, \tau \right] \nonumber\\
&+&\mbox{}  \rm  m_2 \,\, exp \left[-2\sqrt{\frac{2}{3}\pi G
M_{-1} - \frac{\Lambda}{3}} \,\, \tau  \right]
\end{eqnarray}
here  $\rm m_1$ and $\rm m_2$ are integration constants. For
inflation, the following conditions are necessary: $\rm m_1 > m_2$
and $\rm M_{-1} > (\Lambda + 3)/8 \pi G> 0$. This last result
generalize that  found in \cite{inin}, and is the same if $m_2=0$
and $\Lambda=0$ in the gauge $\rm N=1$.
\end{enumerate}
\item{}  $\gamma = -1/3$, the equation (\ref{evolution}) is written
in the following form
\begin{equation}
\rm d\tau = \frac{dx}{\sqrt{b x^2  + a_{-1/3} - \kappa}}
\end{equation}
with solution
\begin{equation}
\rm A(\tau) = \sqrt{\frac{8\pi G M_{-\frac{1}{3}}-3\kappa
}{|\Lambda|} } \quad sinh \left[ \sqrt{\frac{|\Lambda|}{3} } \tau
\right],
\end{equation}
with  $\Lambda < 0$. For inflation, the following conditions are
necessary $\rm M_{-\frac{1}{3}}> 3\kappa/8\pi G >0$,
implying $\kappa =1$ and $|\Lambda| > 3 M_{ -\frac{1}{3}}$.

\item{} $\gamma = - 2/3$. The equation (\ref{evolution}) read as
\begin{equation}
\rm d\tau = \frac{dx}{\sqrt{ b x^2 + a_{-1/3} x - \kappa}}
\label{infla23}
\end{equation}
\end{enumerate} with the solution for $\kappa =-1$ and $\Lambda <0$
is
\begin{eqnarray}
\rm A(\tau) &=& \rm \frac{3}{2|\Lambda|} \left\{
\left(\sqrt{\frac{|\Lambda|}{3}}- \frac{4}{3}\pi G
M_{-\frac{2}{3}}  \right)
e^{\sqrt{\frac{|\Lambda|}{3}}\tau} -  \right.\nonumber\\
& & \mbox{} - \left(\sqrt{\frac{|\Lambda|}{3}}+ \frac{4}{3}\pi G
M_{-\frac{2}{3}} \right)
e^{-\sqrt{\frac{|\Lambda|}{3}}\tau}  -\nonumber\\
 & & \mbox{} - \left. \frac{8}{3}\pi G M_{-\frac{2}{3}} \right\}.
\end{eqnarray}

having an inflationary  behaviour.

\subsection{$\gamma =1/3$,  $\Lambda \not=0$, any $\kappa$ }
In this subcase, (\ref{evolution}) is written as
\begin{equation}
\rm d\tau = \frac{AdA}{\sqrt{\frac{8}{3} \pi G M_{\frac{1}{3}}
-\frac{1}{3}\Lambda A^4- \kappa A^2} }. \label{evolution2}
\end{equation}
with the change of variables $\rm u= A^2$, (\ref{evolution2}) is
\begin{equation}
\tau  =\frac{1}{2}\int_0^{A^2} \frac{du}{\sqrt{\frac{8}{3} \pi G
M_{\frac{1}{3}} -\frac{1}{3}\Lambda u^2- \kappa u} }.
\end{equation}
whose  solutions are, depending on the sign of $\Lambda$

1. $\Lambda > 0$

\begin{eqnarray}
&&\sqrt{\frac{3}{4\Lambda}} \left\{\arcsin
\left[\frac{\frac{2\Lambda A^2}{3} + \kappa}{\sqrt{\kappa^2 +
\frac{32}{9}\pi G \Lambda M_{\frac{1}{3}} }} \right] \right.
\nonumber\\
&&\left. -\arcsin \left[\frac{\kappa}{\sqrt{\kappa^2 +
\frac{32}{9}\pi G \Lambda M_{\frac{1}{3}} }} \right] \right\}
\end{eqnarray}

2.  $\Lambda < 0$
\begin{eqnarray}
&&\sqrt{\frac{3}{4\Lambda}} \left\{ \rm Ln
\left[2\sqrt{-\frac{\Lambda}{3}\left(\frac{8}{3} \pi G
M_{\frac{1}{3}}-\frac{1}{3}\Lambda A^4- \kappa
A^2\right)}\right.\right.\nonumber\\
&&\left. \left. -\frac{2}{3}\Lambda A^2 - \kappa \right] - Ln
\left[ 2\sqrt{-\frac{8}{9} \pi G \Lambda M_{\frac{1}{3}}}- \kappa
\right] \right\}.
\end{eqnarray}

\subsection{$\gamma >0$,  $\Lambda =0$, }
For this subcase, (\ref{evolution}) is given by

\begin{equation}
\rm d\tau = \frac{dA}{\sqrt{\frac{8}{3} \pi G M_\gamma
A^{-(3\gamma +1)}
 - \kappa} },
\label{evolution1}
\end{equation}
and introducing the following conformal transformation $\rm d\tau
= x^{3\gamma +2} dT$, and the change of variable $\rm u= a_\gamma
x^{-(3\gamma +1)} - \kappa$, the scale factor becomes in the
``conformal time'' T,
\begin{equation}
\rm A(T) =  \left[ \frac{a_\gamma(3\gamma +1)^2}{4} T^2 -
\sqrt{-\kappa} (3\gamma +1) T \right]^{-\frac{1}{3\gamma +1}}
\label{scaleT}
\end{equation}
which is valid for $\kappa \le 0$.

When we know the function $\rm A(T)$, we can obtain the
transformation rule between the times $\rm d\tau$ and $\rm dT$,
for instance
\begin{equation}
\rm d \tau = A^{3\gamma +2} dT,
\label{relation}
\end{equation}
thus
\begin{equation}
\rm d\tau = \left[ \mu_\gamma T^2 - \nu_\gamma  T \right]^{
-\frac{3\gamma+ 2}{3\gamma +1}}\, dT \label{transformation}
\end{equation}
where $\rm \mu_\gamma = \frac{a_\gamma(3\gamma +1)^2}{4}$ and $\rm
\nu_\gamma = \sqrt{-\kappa} (3\gamma +1)$.

Now, considering the  flat universe ($\kappa=0$), we can integrate
(\ref{transformation}), but for consistence between  eqs.
(\ref{scaleT}) and (\ref{relation}),  we
 introduce the parameter $\epsilon$ in the sense that when $\tau=0$,
$\epsilon=T$ and $\tau=\tau$, $\epsilon\rightarrow \infty$
\begin{equation}
\int_0^\tau d\tau =\mu^{-\frac{3\gamma+2}{3\gamma +1} }
\int^\epsilon _T \left(x \right)^{
-\frac{2(3\gamma+1)}{3\gamma+1}} dx \label{master}
\end{equation}
After a tedious calculation, we arrive at
\begin{equation}
\rm T = \left[ \mu^{\frac{3\gamma+2}{3\gamma + 1}}
\frac{3(\gamma+1)}{3\gamma + 1} \tau  +
\epsilon^{-\frac{3(\gamma+1)}{3\gamma + 1}}
\right]^{-\frac{3\gamma +1}{3(\gamma+1)}}. \label{TimeT}
\end{equation}
Introducing (\ref{TimeT}) into (\ref{scaleT}) we found the scale
factor in general way
\begin{equation}
\rm A(\tau) = \left[ \sqrt{\mu_\gamma} \frac{3(\gamma +
1)}{3\gamma +1} \tau + \mu^{-\frac{3(\gamma+1)}{2(3\gamma + 1)}}
\epsilon^{-\frac{3(\gamma+1)}{3\gamma + 1}}
\right]^{\frac{2}{3(\gamma+1)}}.
\end{equation}
At this point, we can calculate the behaviour of the scale factor
for some positive values to the parameter $\gamma$ and compare
them with those found in the standard literature,
\begin{enumerate}
\item{} Dust era, $\gamma = 0$, the scale factor become in the
gauge $\rm N=1$
\begin{equation}
\rm A(t)= \left\{ \left[(6\pi G M_0\right]^{\frac{1}{2}} t +
\mu_0^{-\frac{3}{2}} \epsilon^{-3} \right\}^{\frac{2}{3}}.
\end{equation}
\item{} Radiation era,  $\gamma=\frac{1}{3}$, in the gauge $\rm
N=1$
\begin{equation}
\rm A(t) = \left\{ \left[ \frac{32}{3}\pi G M_{\frac{1}{3}}
\right]^{\frac{1}{2}}  t   + \mu_{\frac{1}{3}}^{-1} \epsilon^{-2}
\right\}^{\frac{1}{2}}.
\end{equation}
\item{} Stiff matter, $\gamma = 1$, in the gauge $\rm N=1$
\begin{equation}
\rm A(t) = \left\{ \left[ 24 \pi G M_1 \right]^{\frac{1}{2}} t +
\mu_1^{-\frac{3}{4}} \epsilon^{-\frac{3}{2}}
\right\}^{\frac{1}{3}}.
\end{equation}
Choosing appropriately the parameter $\epsilon \rightarrow
\infty$, we obtain the usual results for the scale factor for the
FRW, with $\kappa=0$ and $\Lambda = 0$ \cite{inin}.
\end{enumerate}

\section{Comments}
The inflationary scenarios for $\rm \gamma <0$ and matter epochs
were considered in the FRW cosmological model. Also, our method is
more general that this employed by Faraoni, because when the
cosmological constant is included in the formalism, the equation
(3.1) in Ref. \cite{faraoni} does not reduce to Riccati equation
and the procedure fails.

\noindent {\bf Acknowledgments}\\
This work was partially supported by PROMEP and Gto. Univ.
projects.

\end{document}